\documentclass[10pt,letterpaper,twocolumn]{article} 

\usepackage{ol2_no_copyright}

\usepackage[draft]{hyperref}
\usepackage{amsmath}
\usepackage{subfigure}
\usepackage{color}

\begin{document}

\twocolumn[ 

\title{All-optical generation of states for ``Encoding a qubit
  in an oscillator''} 


\author{H. M. Vasconcelos,$^{1,*}$ L. Sanz,$^2$ and S. Glancy$^{3}$}

\address{
$^1$Departamento de Engenharia de Teleinform\'atica, Universidade Federal do Cear\'a, Fortaleza, Brazil
\\
$^2$Instituto de F{\'\i}sica, Universidade Federal de Uberl\^andia, Uberl\^andia, Brazil \\
$^3$Mathematical and Computational Sciences Division, National
Institute of Standards and Technology, Boulder, Colorado, USA\\
$^*$Corresponding author: hilma@ufc.br
}

\begin{abstract}
Both discrete and continuous systems can be used to encode quantum
information. Most quantum computation schemes propose encoding qubits
in two-level systems, such as a two-level atom or an electron
spin. Others exploit the use of an infinite-dimensional system, such as
a harmonic oscillator. In ``Encoding a qubit in an
oscillator'' {[Phys.~Rev.~A \textbf{64} 012310 (2001)]}, Gottesman, Kitaev, and Preskill (GKP) combined these
approaches when they proposed a fault-tolerant quantum computation
scheme in which a qubit is encoded in the continuous position and
momentum degrees of freedom of an oscillator. One advantage of this
scheme is that it can be performed by use of relatively simple linear
optical devices, squeezing, and homodyne detection. However, we lack a
practical method to prepare the initial GKP states.  Here we propose
the generation of an approximate GKP state by using superpositions of
optical coherent states (sometimes called ``Schr\"odinger cat
states''), squeezing, linear optical devices, and homodyne detection.
\end{abstract}


 ] 

The Gottesman, Kitaev, and Preskill (GKP) scheme \cite{Gottesman2001}, constitutes a type of linear optical quantum
computer as are other schemes based on the proposals of Knill, Laflamme,
and Milburn \cite{Knill2001} and schemes based on the proposal of 
Ralph {\it et al.}~\cite{Ralph2003}. In the GKP scheme, the qubit is encoded
in the continuous Hilbert space of an oscillator's position and momentum
variables. The GKP scheme is applicable to any type quantum harmonic
oscillator, but we focus on an optical implementation
in traveling modes.  The GKP encoding provides a natural error-correction scheme
to correct errors due to small shifts (applications of the
displacement operator) on the conjugate quadrature variables $x$ and $p$
\cite{Gottesman2001}.

\begin{figure}
\centerline{
\includegraphics[width=5cm]{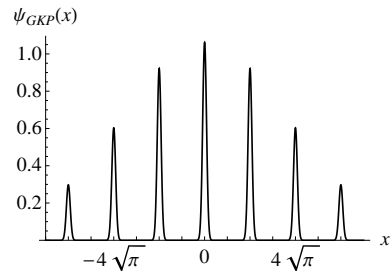}}
  \caption{Approximate GKP state's $x$-quadrature wave function. This
  shows the logical 0 state $\psi_{GKP}(x)$ with $\Delta = k= 0.15$.}
  \label{fig:approxGKP}
\end{figure}

The ideal GKP logical 0 qubit state, $|\bar{0}\rangle$, is defined as
a state whose $x$-quadrature wave function is an infinite series of
delta-function peaks occurring whenever $x = 2\sqrt{\pi}\,s$ for all
integers $s$, while the ideal $|\bar{1}\rangle$ $x$-quadrature wave
function is displaced a distance $\sqrt{\pi}$ from the
$|\bar{0}\rangle$ state. Since these states are unphysical, GKP described approximate
states whose $x$-quadrature wave function is a series of Gaussian
peaks with width $\Delta$, contained in a larger Gaussian envelope of
width $1/k$.  The approximation of $|\bar{0}\rangle$ has the wave function
\begin{equation}
\psi_{GKP}(x)= N\sum_{s=-\infty}^{\infty}e^{-\frac{1}{2}(2sk\sqrt{\pi})^2}e^{-\frac{1}{2} \left(\frac{x-2s\sqrt{\pi}}{\Delta} \right)^2},
\end{equation}
where $N$ is a normalization factor. If $\Delta$ and $k$ are small,
then $\psi_{GKP}$ will better approximate an ideal GKP state, and the wave
function will have many sharp Gaussian peaks contained in a wide
envelope. We can think of the deviation from an ideal GKP state as
corresponding to nonzero probability that the state has suffered from
errors causing displacement in the $x$ or $p$ variables.  If all
displacements are smaller than $\sqrt{\pi}/6$, then the errors will
not increase during the error correction protocol.  GKP states with
$\Delta<0.15$ and $k<0.15$ will have a probability greater than $0.99$ of being free
of shift errors larger than $\sqrt{\pi}/6$\cite{Glancy2006}. 
Fig.~\ref{fig:approxGKP} shows an example of an approximate GKP
state's $x$-quadrature wave function.

Preparing GKP states is a difficult task, and no experiment has yet
demonstrated preparation of such states. GKP proposed preparing these
states by coupling an optical mode to an oscillating mirror
in~\cite{Gottesman2001}. Another proposal was made by Travaglione and
Milburn in~\cite{Travaglione2002}, where the qubit states are prepared
in the oscillatory motion of a trapped ion rather than the photons in
an optical mode.  Pirandola {\em et al.}~\cite{Pirandola2004}
discusses the preparation of optical GKP states by use of a two mode Kerr
interaction followed by a homodyne measurement of one of the
modes. The same authors describe two proposals for generating GKP states
in the position and momentum of an atom by using a cavity mediated
interaction with light in~\cite{Pirandola2006,Pirandola2006-1}. 

Here we propose the generation of an approximate GKP state by using
superpositions of optical coherent states (``cat states''), linear
optical devices, squeezing, and homodyne detection. The basic idea
is: first, prepare two cat states (each of which contains two Gaussian
peaks in its $x$-quadrature wave function), squeeze both cats (to
reduce the width of the Gaussian peaks), interfere them at a beam
splitter, then perform homodyne detection on one of the beam
splitter's output ports. Depending on the measurement result, we
will find an approximate GKP state (with three Gaussian peaks) in
the beam splitter's other output port.  This procedure can be repeated
to produce states with larger numbers of Gaussian peaks.

A cat state is a superposition of coherent states such as:
\begin{equation}
|\psi_{\textrm{cat}}\left(\alpha\right) \rangle = \frac{|-\alpha \rangle+ |\alpha \rangle}{\sqrt{2(1+e^{-2\alpha^2})}},
\label{eq:cat}
\end{equation}
where $\alpha$ is the amplitude of the coherent state, which may be
complex, but we assume it is real below.  Several experimental proposals to create
cat states are reviewed in~\cite{Glancy2008}. Cat states of this form
have been created in several experiments with $|\alpha|$ up to 1.75
and fidelities of 0.6 to 0.7\cite{Ourjumstev2006,
Neergaard-Nielsen2006, Wakui2007, Ourjoumtsev2007, Takahashi2008,
Gerrits2010}. 

Cat states' $x$-quadrature wave functions are superpositions of
Gaussian peaks. To simplify notation, we denote a Gaussian with
\begin{equation}
G(x,V,\mu)=e^{\frac{-(x-\mu)^2}{2V}}.
\end{equation}
These Gaussians represent wave functions, not probability
distributions, so the unnormalized vacuum state is
$G(x,1,0)$. Suppose two modes, labeled 1 and 2, contain states with unnormalized wave functions
$G(x_1,V,\mu_1)$ and $G(x_2,V,\mu_2)$.  Modes 1 and 2 meet at a beam splitter
with transmissivity $1/2$, which performs the transformation:
\begin{eqnarray}
x_1 \rightarrow \frac{1}{\sqrt{2}}\left(x_1+x_2\right) \nonumber \\
x_2 \rightarrow \frac{1}{\sqrt{2}}\left(x_1-x_2\right).
\end{eqnarray}
After the beam splitter, we use a homodyne detector to measure mode
2's $p$-quadrature. In the case that the measurement result is
$p_2=0$, this entire procedure produces
the transformation
\begin{equation}
G(x_1,V,\mu_1)G(x_2,V,\mu_2)\rightarrow\sqrt{V} G(x_1,V,\frac{\mu_1+\mu_2}{\sqrt{2}}).
\label{beamSplitMeasure}
\end{equation}

We can write Eq.~(\ref{eq:cat}) in the $x$-quadrature basis as:
\begin{equation}
\tilde{\psi}_{\textrm{cat}}(x,\alpha)=G(x,1,-\sqrt{2}\alpha)+G(x,1,\sqrt{2}\alpha),
\end{equation}
where we use the tilde to signal that the state is not normalized.  We
now squeeze this state by an amount $\zeta$, obtaining
\begin{eqnarray}
\tilde{\psi}_{\textrm{sqcat}}(x,\alpha,\zeta) & = & G(x,e^{-2\zeta},-\sqrt{2}\alpha
e^{-\zeta}) \nonumber \\
 & & + G(x,e^{-2\zeta},\sqrt{2}\alpha e^{-\zeta}).
\end{eqnarray}
We will choose the cat state amplitude to be
$\alpha=\sqrt{2}^{m-1}\sqrt{\pi} e^\zeta$, where $m$ is an integer greater
or equal to $1$, which we will later use to count iterations of our scheme. With these choices,
\begin{equation}
\begin{split}
\tilde{\psi}_{\textrm{sqcat}}(x,\sqrt{2}^{m-1}\sqrt{\pi} e^\zeta, \zeta)=\,&G(x,e^{-2\zeta},-\sqrt{2}^m\sqrt{\pi})\, + \\
& G(x,e^{-2\zeta},\sqrt{2}^m\sqrt{\pi}).
\end{split}
\end{equation}
Suppose we have two copies of this squeezed cat in modes 1 and 2.  They
meet at a beam splitter with transmissivity $1/2$, and the
$p$-quadrature of mode 2 is measured to be $p_2=0$.  If we choose $m=1$,
the resulting unnormalized state of mode 1 is
\begin{equation}
\begin{split}
\tilde{\beta}(x_1,\zeta,m=1)=\,G(x_1,e^{-2\zeta},-2\sqrt{\pi})\,+ &\\
2G(x_1,e^{-2\zeta},0)\,+ G(x_1,e^{-2\zeta},2\sqrt{\pi}),&
\end{split}
\end{equation}
which we call the first binomial state and it is similar to an
approximate GKP logical qubit 0, except only the central three peaks
are present.  The width of those peaks is controlled by the amount of
squeezing $\zeta$ applied to the original cat states.

Consider the order $m$ binomial state given by
\begin{equation}
\tilde{\beta}(x,\zeta,m)=\sum_{n=0}^{2^m}\binom{2^m}{n}G\left[x,e^{-2\zeta},2\sqrt{\pi}\left(n-2^{m-1}\right)\right].
\label{eq:beta}
\end{equation}
This state is a series of Gaussian peaks separated by $2\sqrt{\pi}$
along the $x$-quadrature axis, and the amplitudes of the peaks are
given by the ($2^m$)\textsuperscript{th} row of Pascal's Triangle (where row 0 contains
only 1).  We will show that given two copies of the order $m$ binomial
state, one can make the $m+1$ order binomial state.  We begin with the
state of modes 1 and 2:
\begin{equation}
\begin{split}
\tilde{\beta}(x_1,\zeta,m)\tilde{\beta}(x_2,\zeta,m) =
\sum_{n_1=0}^{2^m}\sum_{n_2=0}^{2^m}\binom{2^m}{n_1}\binom{2^m}{n_2}
\\ 
 \times G\left(x_1,e^{-2\zeta},2\sqrt{\pi}(n_1-2^{m-1})\right)
 \\
 \times G\left(x_2,e^{-2\zeta},2\sqrt{\pi}(n_2-2^{m-1})\right)
\end{split}
\end{equation}
These two modes meet in a beam splitter of transmissivity $1/2$, and we
measure the $p$-quadrature, obtaining the result $p_2=0$.  The new state is
given by applying Eq.~(\ref{beamSplitMeasure}) to
$\tilde{\beta}(x_1,\zeta,m)\,\tilde{\beta}(x_2,\zeta,m)$.  The result is
\begin{eqnarray}
\sum_{n_1=0}^{2^m}\sum_{n_2=0}^{2^m}\binom{2^m}{n_1}\binom{2^m}{n_2}\,\times \nonumber \\
G\left(x_1,e^{-2\zeta},\sqrt{2}\sqrt{\pi}(n_1+n_2-2^m)\right).
\end{eqnarray}
After a little algebra and application of Vandermont's identity, we obtain
\begin{eqnarray}
\sum_{q=0}^{2^{m+1}}\binom{2^{m+1}}{q} \,
G\left(x_1,e^{-2\zeta},\sqrt{2}\sqrt{\pi}(q-2^m)\right).
\label{eq:iteraction}
\end{eqnarray}
This state is equivalent to $\beta(x_1,\zeta,m+1)$, except that the
Gaussian peaks are separated by only $\sqrt{2}\sqrt{\pi}$ rather than
$2\sqrt{\pi}$.  We can compensate for this shrinking of the spacing
between the Gaussian peaks, if we begin the
procedure with two binomial states with spacing of
$2\sqrt{2}\sqrt{\pi}$. As $m$ increases, these binomial states
will approach the shape of a series of Gaussian peaks in a Gaussian
envelope, like $\psi_{GKP}(x)$.

\begin{figure}
   \centering
     \includegraphics[width=4.1cm]{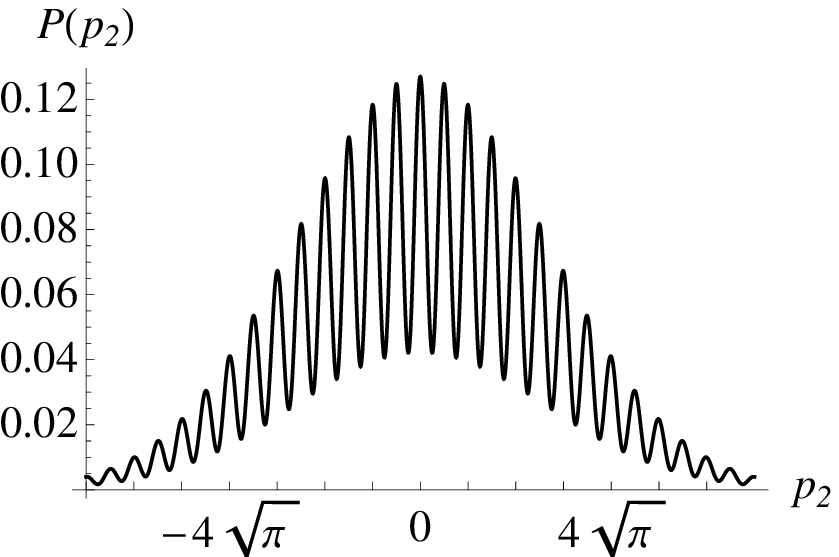}
     \includegraphics[width=4.1cm]{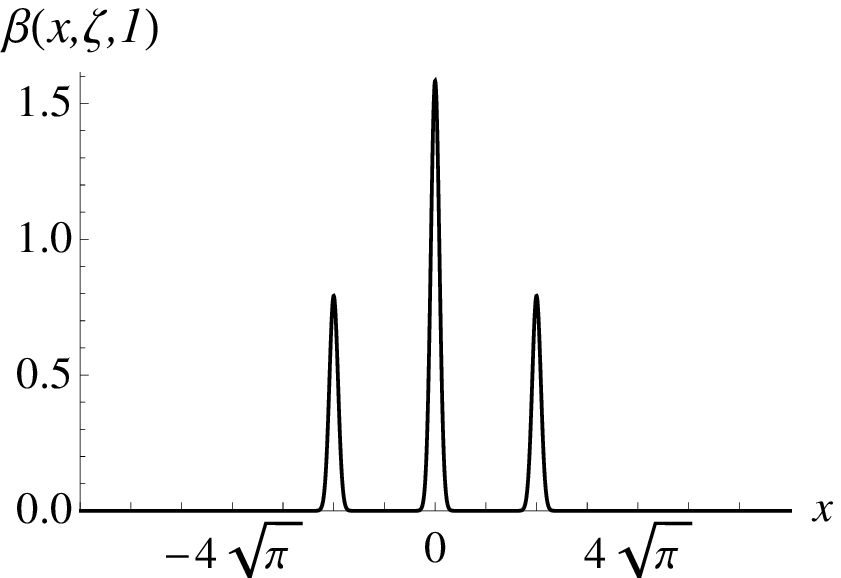}
     \caption{Left: Probability for measuring $p_2=R$ as function of
     $R$. Right: $x$-quadrature wave function for the state $\beta(x,\zeta,1)$.
     In both cases $\alpha = \sqrt{\pi} e^{\zeta}$, and $\zeta =
     1.9$ is chosen to produce Gaussian peaks with the same width as
     shown in Fig.~\ref{fig:approxGKP}. This $\zeta$ is equivalent
     to -16 dB of quadrature noise power reduction in conventional
     squeezing experiments.}
     \label{fig:estado0-3p}
\end{figure}

\begin{figure}
   \centering
\includegraphics[width=5cm]{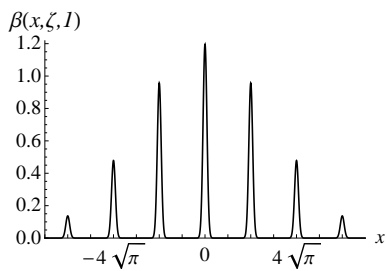}
\caption{Wave function of binomial state $\beta(x,\zeta,3)$, a closer
  approximation of the logical 0 GKP state.  Again here $\zeta=1.9$.
  Creating this state would require at least eight cat states, each with
  $\alpha=2\sqrt{\pi}e^\zeta$.  The $\beta(x,\zeta,3)$ state also has
  small peaks at $\pm8\sqrt{\pi}$, which are not visible here.}
     \label{fig:estado0-9p}
\end{figure}

In Fig.~\ref{fig:estado0-3p}, we plot the probability for measuring a
certain value $p_2=R$ as a function of $R$ when making the $m=1$
binomial state from two cat states. If we measure $R=0$, we obtain the
wave function as shown in Fig.~\ref{fig:estado0-3p}, whose Gaussian
peaks' widths are determined by the degree of squeezing applied to the
initial cat states, and whose heights are proportional to the second
row of Pascal's triangle $(1,2,1)$, as given by
Eq.~(\ref{eq:beta}). In Fig.~\ref{fig:estado0-9p} we show the $m=3$
binomial state.

Creating the order $m$ binomial state requires a minimum of $2^m$
cats, but the true number may be much larger, because we require that
$p_2=0$ at each measurement event. However, initial investigations
indicate that some cases in which $p_2 \neq 0$ can be recovered by
applying a $p$-quadrature displacement whose size depends on the
measurement result.  We plan to explore this further in a future work.
Creating high quality states will require larger squeezing $\zeta$ and
higher order $m$ binomial states.  We also plan to investigate the
performance of these binomial states in the GKP encoding scheme as a
function of $\zeta$ and $m$.

In this paper we began construction of GKP with a source of cat
states.  The most popular method to make cat states is by subtracting
photon(s) from a squeezed vacuum state.  It may be possible to alter
the photon subtraction scheme to benefit our method to make GKP
states.

Although our scheme is built of apparently simple, well understood
 optical operations, it will  be  difficult to
achieve in an experiment.  We expect that matching the transverse and
longitudinal shapes of all of the optical modes, especially during the
squeezing stage~\cite{Wasilewski2006,Cohen2008}, will be very
difficult.

We thank Adam Meier, Yanbao Zhang, and Emanuel Knill for helpful
discussion and comments.  H. M. Vasconcelos thanks the PNPD program
(CAPES) and FUNCAP for financial support. L. Sanz thanks INCT-IQ for
financial support.  This paper is a contribution by the National
Institute of Standards and Technology of the United States of America
and not subject to USA copyright.

\end{document}